# Focusing Viral Risk Ranking Tool on Prediction

Katherine Budeski[1], MMSc, MSc & Marc Lipsitch[1], DPhil
[1] Department of Epidemiology, Harvard T.H. Chan School of Public Health

## Abstract

Preparing to rapidly respond to emerging infectious diseases is becoming ever more critical. *SpillOver: Viral Risk Ranking* is an open-source tool developed to evaluate novel wildlife-origin viruses for their risk of spillover from animals to humans and their risk of spreading in human populations. However, several of the factors used in the risk assessment are dependent on evidence of previous zoonotic spillover and/or sustained transmission in humans. Therefore, we performed a reanalysis of the *Ranking Comparison* after removing eight factors that require post-spillover knowledge and compared the adjusted risk rankings to the originals. The top 10 viruses as ranked by their adjusted scores also had very high original scores. However, the predictive power of the tool for whether a virus was a human virus or not as classified in the Spillover database deteriorated when these eight factors were removed. The area under the receiver operating characteristic curves (AUROC) for the original score, 0.94, decreased to 0.73 for the adjusted scores. Furthermore, we compared the mean and standard deviation of the human and non-human viruses at the factor level. Most of the excluded spillover-dependent factors had dissimilar means between the human and non-human virus groups compared to the non-spillover dependent factors, which frequently demonstrated similar means between the two groups with some exceptions. We concluded that the original formulation of the tool depended heavily on spillover-dependent factors to "predict" the risk of zoonotic spillover for a novel virus. Future iterations of the tool should take into consideration other non-spillover dependent factors and omit those that are spillover-dependent to ensure the tool is fit for purpose.

# Introduction

Following the COVID-19 pandemic, several efforts to prevent, detect, and rapidly respond to emerging infectious diseases were launched, including the 100 Days Mission, which aims for global rollout of a vaccine to a novel pathogen within 100 days after detection.[1] Given that approximately 60% of emerging infectious diseases are zoonoses and a majority originate in wildlife,[2] virus prospecting–or the discovery and categorization of new zoonotic viruses prior to spillover–has been a focus of major efforts such as the Global Virome Project,[3] PREDICT,[4] and DEEP VZN, which was terminated in July 2023.[5]

The *SpillOver: Viral Risk Ranking* tool (Spillover tool) presents a publicly available and interactive platform "for policy makers, scientists and the general public to assess the likelihood that a wildlife virus will spillover and spread in humans".[6] The results of the ranking exercise, published by Grange et al., show that the top 12 viruses identified had indeed spilled over at the time of publication, a result described as "[v]alidating the risk assessment".[6] Inputs to the Spillover tool can evolve over time. As of August 13, 2024, the top 10 viruses ranked by the tool are all known to infect humans: SARS-CoV-2, Lassa, Ebola, Seoul, Nipah, Hepatitis E, Marburg, Simian Immunodeficiency, Rabies, and Lymphocytic Choriomeningitis viruses. We note, however, that several of the factors used to quantify a virus' risk of spillover include prior evidence of zoonotic spillover itself or widespread transmission in humans. We therefore performed a reanalysis of the Spillover tool, taking the data as downloaded from https://spillover.global/ranking-comparison/ but eliminating from the scoring those factors that could only be satisfied once a virus of interest has spilled over and/or caused widespread transmission in humans.[6] Adjusted scores excluding these factors were calculated, normalized, and compared to the original scores and ranks while maintaining the original weights.

# Results

Eight factors were identified as partially or fully dependent on evidence of spillover of a virus of interest (Table 1). Adjusted scores were recalculated after excluding these factors, and both the original scores and adjusted scores were normalized relative to the highest score by each set of factors.

*Table 1. Risk factors as well as a short explanation of the risk factors as written in the SpillOver: Viral Risk Ranking tool whose scores were excluded from the recalculations.*

| Risk Factor - SpillOver Format | Short Explanation of Risk Factor |
|---|---|
| Animal to human transmission | Known ability of the virus to transmit between wildlife and people (A zoonotic virus). |
| Human to human transmission | Known ability of the virus to transmit between people. |

| Duration of infection in humans | Whether the virus species is known to chronically (> 4 weeks) or acutely (< 4 weeks) infect people. |
|---|---|
| Viral infectivity in humans | Detection of a virus species in a human. |
| Proportion of viruses known to infect humans in the viral family | Proportion of virus species within the viral family that have been detected in a human. |
| Epidemicity of the virus | Whether the virus species has been implicated in pandemic or epidemic/outbreak events in humans or animals (mammals, birds, reptiles, and amphibians). |
| Pandemic virus | Whether the virus has caused a pandemic in humans. |
| Proportion of known human pathogens in the viral family | Proportion of virus species within a viral family that are known to cause disease in people. |

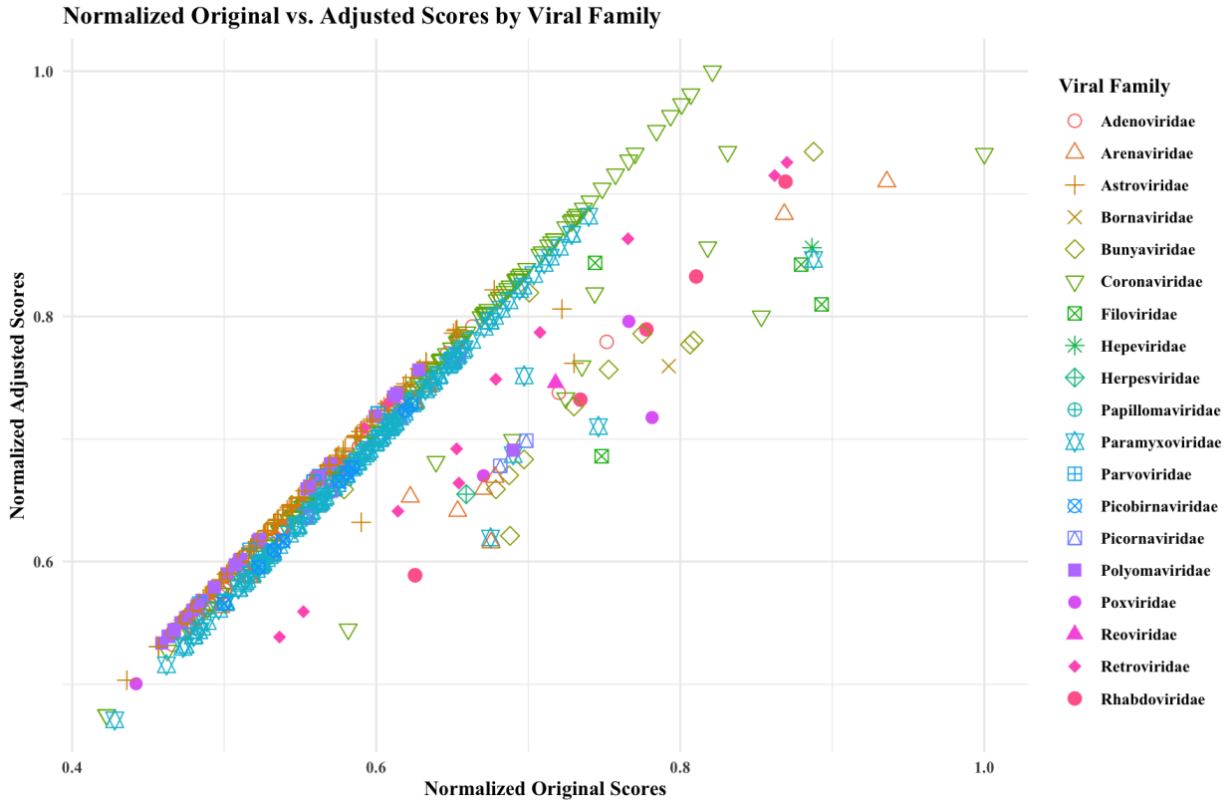

*Figure 1. Normalized adjusted vs. normalized original risk scores using factors and their weights from the Spillover tool.* Viruses are coded by shape and color to represent their viral families. Viruses with missing values for their viral family were excluded from the plot.

**Figure 1** displays the normalized adjusted (y-axis) and original (x-axis) risk scores for every virus listed within the *Spillover Virus Ranking* before May 22, 2024. A strong near-diagonal relationship demonstrates that many viruses' scores were affected modestly by the removal of the spillover-dependent factors, while the concentration of points to the right of the diagonal represents those viruses whose normalized scores went down substantially after the removal of the spillover-dependent factors.

The top 10 ranked viruses using the adjusted scores are: Rousettus bat coronavirus HKU9 (73.9), Murine coronavirus (72.5), Chaerephon bat coronavirus/Kenya/KY22/2006 (72.0), Coronavirus PREDICT CoV-35 (71.2), Longquan Aa mouse coronavirus (70.3), Seoul virus (69.1), Coronavirus 229E (Bat strain) (69.1), Eidolon bat coronavirus/Kenya/KY24/2006 (69.0), Severe acute respiratory syndrome coronavirus 2 (69.0), and Coronavirus PREDICT CoV-24 (68.6). The original ranks of these viruses using the original score ranged from 1 (SARS-CoV-2) to 29 (Coronavirus PREDICT CoV-24), indicating that the top 10 viruses by the adjusted scores had also scored very high on their original scores.

a)

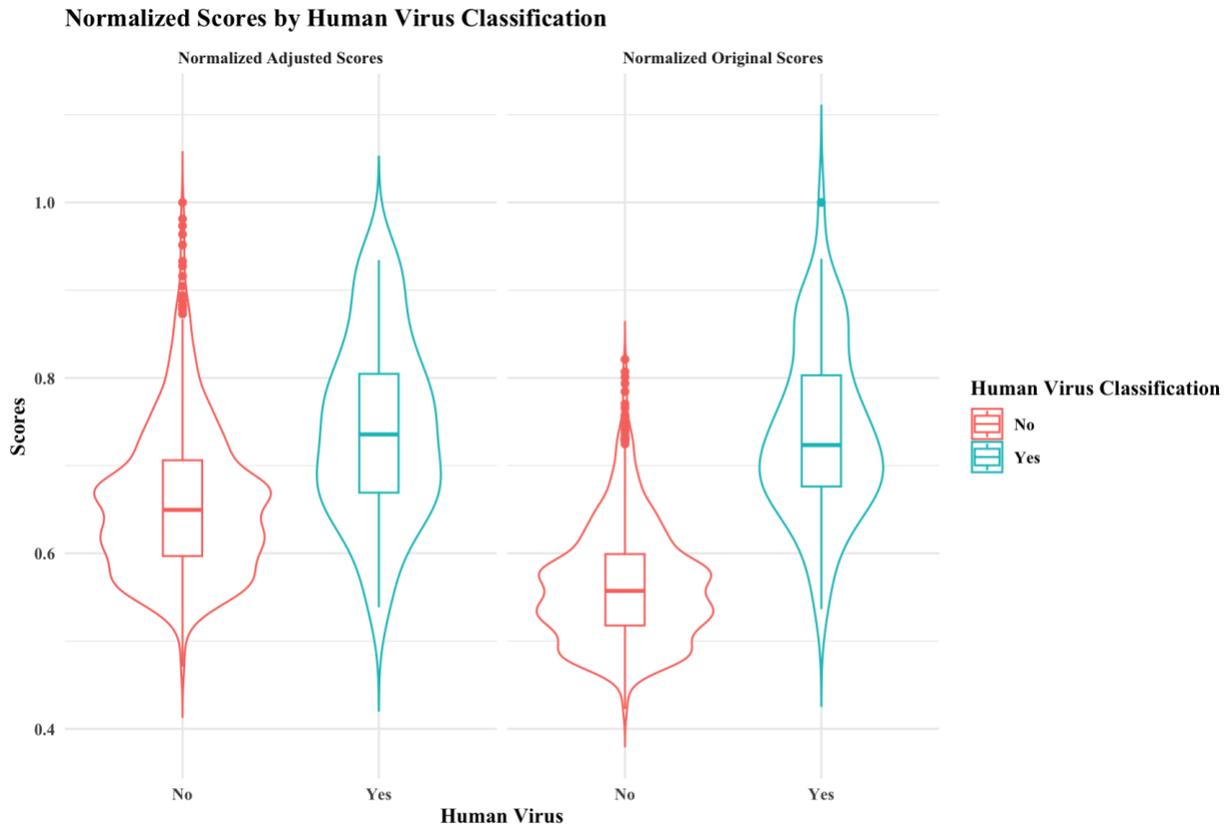

b)

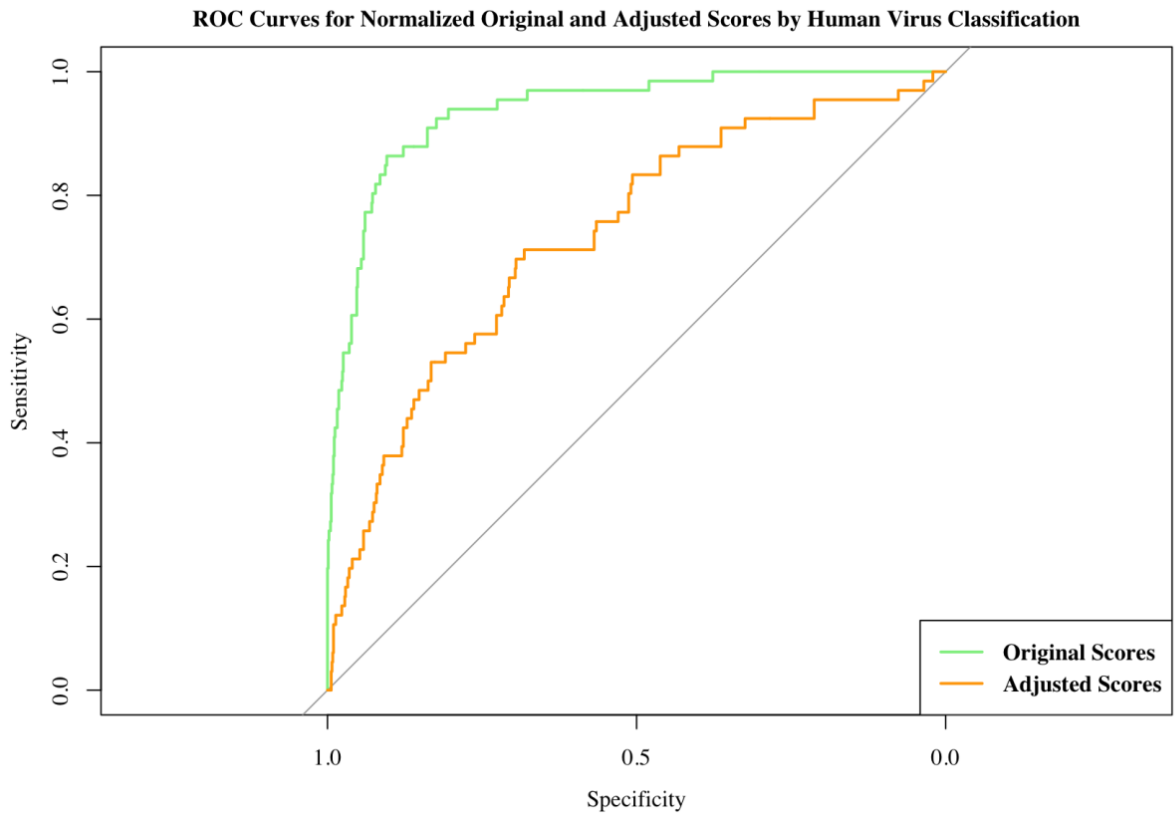

*Figure 2. (a) Distributions of normalized scores by whether the Spillover tool dataset classifies the virus as a "Human Virus" comparing normalized adjusted (left) vs. original (right) risk scores. (b) Receiver Operating Characteristic curve for the normalized adjusted and original risk scores.* All viruses listed before May 22, 2024, were included.

**Figure 2 (a)** shows normalized adjusted (left) and original (right) risk scores comparing viruses classified as human viruses and non-human viruses stratified by a yes/no category entitled "Human Virus?" within the Spillover tool that is listed for every virus in the online database. Separation is less impressive in the adjusted scores, reflecting the removal of spillover-dependent factors included in the original risk scores that could be satisfied only by human viruses. Quantifying discriminatory power by the area under the receiver operating characteristic curves (AUROC) shows an AUROC of 0.94 for the original risk scores versus 0.73 for the adjusted risk scores (Fig. 2(b)). This indicates that while the adjusted risk scores still have some predictive power, removal of spillover-dependent factors makes them considerably less strongly associated with observation of human infections than the original risk scores.

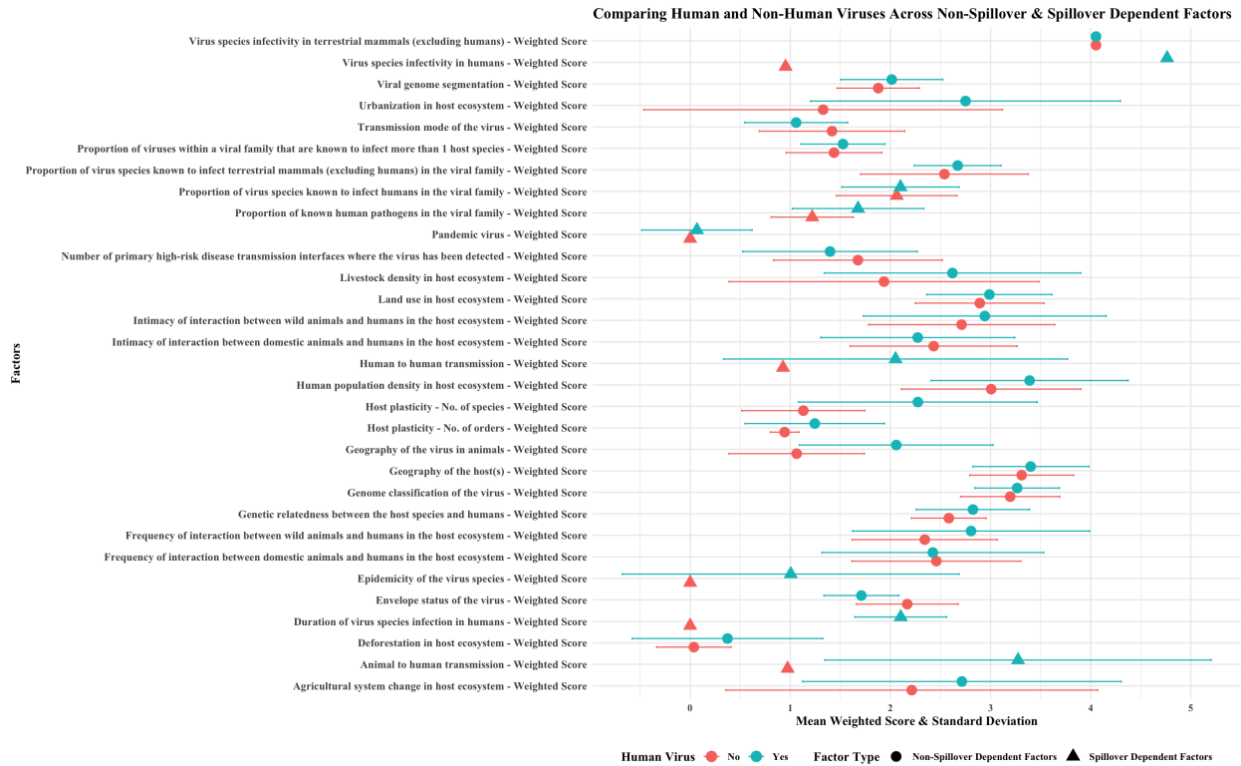

*Figure 3. Mean weighted scores and standard deviations for the human (blue) and non-human (red) viruses for all factors included in the Spillover tool's risk score.* Spillover-dependent factors are indicated with a triangle while non-spillover dependent factors are indicated with a circle. All viruses listed before May 22, 2024, were included.

**Figure 3** shows the mean and standard deviation of the human and non-human viruses by individual factor and highlights how spillover-dependent factors contribute to the Spillover tool's ability to "predict" that a novel virus will spillover and transmit among humans. Interestingly, there were other factors that were not obviously spillover-dependent that contributed to the distinction between the human and non-human virus groups. Such factors include "urbanization in the host ecosystem," "host plasticity–number of species," and "geography of the virus in animals."

# Discussion

As a detected virus spreads, we learn more about its pathogenicity and transmissibility. In their 2021 paper, published the year after SARS-CoV-2 started a global pandemic, Grange et al. noted that SARS-CoV-2 was originally ranked in the *Ranking Comparison* second to Lassa virus.[6] The authors credit the lower ranking to the lack of data available for SARS-CoV-2. Notably, the version of the tool now available online ranks SARS-CoV-2 as the #1 threat, providing further evidence of how the addition of post-spillover data affects the original scoring system to favor pathogens with a known spillover event.

The declared goal of the Spillover tool is to "systematically evaluate novel wildlife-origin viruses in terms of their zoonotic spillover and spread potential".[6] A tool for that purpose should use only factors that are available and relevant for novel wildlife-origin viruses, and not those factors which are available only after spillover and widespread transmission has been observed. Validating the tool while employing post-spillover data and finding that the tool's output matches observed spillovers risks generating circularity.

Our analysis, which maintains the structure of the assessment while removing factors that partially or fully rely on post-spillover data, shows that the top 10 threats identified by the tool change, replacing 8 of the original top 10 with 7 coronaviruses and 1 hantavirus that are not yet known to have spilled over into humans, replacing one with a coronavirus (229E) that has occurred in humans,[7] and preserving SARS-CoV-2, which moves from the first to the seventh rank. While the adjusted scoring system maintains predictive ability for viruses that are known to have spilled over, it is unsurprisingly less predictive than the scoring system that incorporates knowledge of spillover among the predictors.

The recalculated scoring system is less discriminatory between viruses that have been known to spill over and those that have not to date, which we visualized both by comparing the score distributions between known human viruses and those that are not known to infect humans to date, and by comparing the ROC curves of the two scoring systems in distinguishing these two categories. The overlap of distributions is greater, and the ROC curves correspondingly closer to the diagonal, for the recalculated scores without spillover-dependent factors (Fig. 2).

Furthermore, some of the excluded spillover-dependent factors ranked at the top of the list of factors assessing spillover risk as determined by expert opinion and were therefore given a higher weight (*Risk Factor Influence*) in the final score.[6] The top three factors by spillover risk determined by expert opinion were all excluded in our assessment, including "animal to human transmission," "human to human transmission," and "virus species infectivity in humans".[6] By observing the mean and standard deviation for each individual factor by human and non-human viruses we were also able to observe some factors that were not obviously spillover-dependent yet did contribute to the distinction between the human and non-human virus groups (Fig. 3).

There are interesting practical and conceptual questions about how to validate a tool designed to predict relatively infrequent future events. Paradoxically, a tool that perfectly predicts past events (which would have an AUROC=1) may be less than ideal for practical predictive tasks, since it would say that the most likely threats had already spilled over. Given that new spillover events do happen, and that some stochasticity is involved in which viruses are observed to spillover, and when, an ideal evaluation might evaluate factors over time and calculate the distribution of scores at various time points when spillover events happened, asking whether those that spilled over in the past were rated highly likely to spill over with data available at that time. This may be close to impossible in practice. Plowright et al. have also noted the importance of multiple factors aligning to permit spillover, suggesting that an additive model of individual factors may be less useful than one with functional flexibility to account for interactions of multiple factors or requirements for key individual factors.[8] We encourage further methodological refinement to better evaluate such predictive systems, rigorous adherence to using criteria known prior to spillover, and further exploration of the predictive power of the factors that are not spillover-dependent but are nonetheless relevant for viruses that have spilled over.

## Materials and Methods

The *Virus Risk Ranking Assessment* consisted, as described by Grange et al. of three components: (1) identifying risk factors, (2) seeking expert opinions, and (3) conducting the risk calculations. Overall, 31 factors were included in their final "comparative risk score".[6] These 31 factors were the subject of our evaluation.

First, all factors that would score on a new criterion if the virus in question had a known spillover event and/or shown widespread transmission within humans were identified. Eight factors were flagged to have met this criterion (see Table 1).

Once these factors were identified the comparative risk scores and risk rankings were then recalculated. The adjusted risk score included all the factors except the spillover-dependent factors. The original risk score matches that of the Spillover tool and includes all 31 risk factors. The expert weights were maintained for each factor as determined by the *Risk Factor Influence* described by Grange et al.[6] The adjusted and original risk scores were then normalized to the highest score in each set of scores.

The data used for the analysis was downloaded as a CSV file on May 22, 2024, from https://spillover.global/ranking-comparison/. Therefore, the analysis only contains viruses submitted to the platform before this date. The original publicly available CSV was organized using a Jupyter notebook[9] run in Google Colab. Pandas was the main library used for data manipulation.[10] All further data visualizations and analysis were run using a R script[11] in R Studio Version 4.2.3.[12] The tidyverse and dplyr packages were used for data manipulation and visualization.[13] The Receiver Operating Characteristic curves were created using the pROC package.[14] ChatGPT assisted the initial drafting of both the Python and R scripts used for this analysis.[15] All draft code was modified, reviewed, and expanded upon to meet the specific needs of this analysis. All final code has been tested and validated to ensure that it accurately meets the objectives of this analysis.

## Data availability

The data used for this analysis is publicly available and can be downloaded at https://spillover.global/ranking-comparison/.

## Associated Code

All associated code used for this analysis can be found at https://github.com/c2-d2/Spillover-Evaluation. All figures can be reproduced using the instructions and code provided in the repository.

## Acknowledgements

Funding: M.L. thanks the VK Fund for CCDD, Longview Longtermism Fund, Open Philanthropy, and the DALHAP fund for supporting this work.

## Author Contributions

M.L. and K.B. conceived of the study. K.B. performed the data analysis. M.L. and K.B. interpreted the results. K.B. drafted the manuscript. M.L. reviewed and edited the manuscript.

## Competing Interest Statement

The authors declare no competing interests.